\documentclass[useAMS,usenatbib]{mn2e}

\newcommand{\be} {\begin{equation}}

\newcommand{\ee} {\end{equation}}

\newcommand{\psr}{PSR\,B1931+24\,}

\newcommand{\bc}{\begin{center}}
\newcommand{\ec}{\end{center}}
\def\ltsima{$\; \buildrel < \over \sim \;$}
\def\lsim{\lower.5ex\hbox{\ltsima}} 
\def\loe{\lower.5ex\hbox{\ltsima}}
\def\gtsima{$\; \buildrel > \over \sim \;$}
\def\gsim{\lower.5ex\hbox{\gtsima}}
\def\goe{\lower.5ex\hbox{\gtsima}}

\title[On the nature of the intermittent pulsar PSR\,B1931+24]{On the nature of the intermittent pulsar PSR\,B1931+24}

\author[Rea et al.]{N. Rea$^{1,2}$, M. Kramer$^{3}$, L. Stella$^{4}$, P.G. Jonker$^{2,5}$, C.G. Bassa$^{2,6}$, P.J. Groot$^{6}$, \newauthor G.L. Israel$^{4}$, M. M\'endez$^{7}$, A. Possenti$^{8}$, A. Lyne$^{3}$ \\
$^{1}$ University of Amsterdam, Astronomical Institute ``Anton Pannekoek'', Kruislaan 403, 1098~SJ, Amsterdam, The Netherlands\\
$^{2}$ SRON Netherlands Institute for Space Research, Sorbonnelaan 2, 3584~CA, Utrecht, The Netherlands \\
$^{3}$ Jodrell Bank Centre for Astrophysics, University of Manchester, Alan Turing Building, M13~9PL, Manchester, UK\\ 
$^{4}$ INAF--Astronomical Observatory of Rome, via Frascati 33, 00040, Monteporzio Catone (RM), Italy \\
$^{5}$ Harvard-Smithsonian Center for Astrophysics, 60 Garden Street, Cambridge, MA 02138, USA \\
$^{6}$ Department of Astrophysics, Radboud University Nijmegen, PO Box 9010, NL-6500 GL Nijmegen, The Netherlands \\
$^{7}$ Kapteyn Astronomical Institute, Groningen University, 9700 AV, Groningen, The Netherlands \\
$^{8}$ INAF--Astronomical Observatory of Cagliari, Poggio dei Pini, Strada 54, 09012 Capoterra (CA), Italy 
}

\input psfig.sty

\begin{document}

\pagerange{\pageref{firstpage}--\pageref{lastpage}} \pubyear{2006}

\maketitle

\label{firstpage}

\begin{abstract}

\psr\, is the first intermittent radio pulsar discovered to date, characterized by a 0.8\,s pulsation which turns on and off quasi-periodically every $\sim$35\,days, with a duty cycle of $\sim$10\%. We present here X--ray and optical observations of \psr\, performed with the {\it Chandra X-ray Observatory} and  {\it Isaac Newton Telescope}, respectively. Simultaneous monitoring from the {\it Jodrell Bank Observatory} showed that this intermittent pulsar was in the radio--on phase during our observations.  We do not find
any X--ray or optical counterpart to \psr, translating into an upper limit of $\sim 2\times10^{31}$\,erg\,s$^{-1}$ on the X--ray luminosity , and of $g^{\prime} > 22.6$ on the optical magnitude . If the pulsar is isolated, these limits cannot constrain the dim X-ray and optical emission expected for a pulsar of that age ($\sim$1.6\,Myr).  We discuss the possibility that  the quasi--periodic intermittent behavior of \psr\, is due to the presence of a low mass companion star or gaseous planet, tight with the pulsar in an eccentric orbit. In order to constrain the parameters of this putative binary system we re-analysed the pulsar radio timing residuals and we found that (if indeed hosted in a binary system), \psr\, should have a very low mass companion and an orbit of low inclination.

\end{abstract}


\section{Introduction}

A long term radio monitoring study of the $\sim813$\,ms radio pulsar \psr\, (Stokes et al.~1985; Hobbs et al.~2004), revealed the peculiar intermittent behavior of this pulsar (Kramer et al.~2006).
\psr\, is (so far) a unique system: it shows an active radio emission phase lasting between 5--10 days
(radio--on phase, hereafter), which suddenly (in less than 10\,s) switches off, and the pulsar remains undetectable for the following 25--35 days (radio--off phase, hereafter). This pattern repeats
quasi-periodically, and has been monitored for the past 7 years
(Kramer et al.~2006). Another peculiar property of \psr\, is its
spin--down behavior. Remarkably, the pulsar rotation slows down considerably 
faster (by about 50\%) when the pulsar is in the radio--on phase, with a
frequency derivative changing from $\dot{\nu}_{{\rm on}} =
-(16.30\pm0.04)\times10^{-15}$\,s$^{-2}$ to $\dot{\nu}_{{\rm off}}
= -(10.80\pm0.02)\times10^{-15}$\,s$^{-2}$ across the two phases.

From the radio observations, typical pulsar characteristics were
derived (see Tab.\,1 in Kramer et al.~2006): an estimate of
the dipolar magnetic field ($B\sim2.6\times10^{12}$\,G), the
characteristic age ($\tau_c \sim 1.6$\,Myr), and the dispersion measure ($DM = 106.03\pm0.06$\,cm$^{-3}$\,pc). The latter gave a rough
estimate of the pulsar distance of $\sim4.6$\,kpc using the NE2001
model for the free electron distribution in the Galaxy (Cordes \&
Lazio 2002). Kramer et al.~(2006) interpreted the peculiar spin-down
properties of this pulsar in terms of a transient plasma flow in the
magnetosphere, causing the quenching and re-ignition of the radio
emission. In particular, an increase of the magnetospheric current
flow during the radio-on phase, can provide an additional braking torque on the neutron star.  From the observed variations in the pulsar spin-down, these authors showed that the pulsar magnetospheric electron density was
consistent with the Goldreich-Julian density (Goldreich \&
Julian~1969). This interpretation can successfully explain the
variable neutron star torque, but still fails to explain what causes
the changes in the magnetospheric plasma flow, especially if in a
quasi--periodic fashion. To date, this periodicity of the radio--on
and radio--off recurrence is difficult to explain in any scenario
considering an isolated pulsar. As Kramer et al.~(2006) pointed out,
the short shut-off time of less than 10 seconds, and the rather stable
pulse profile, rule out possible scenarios like precession of the
neutron star. On the other hand, the long radio--off phase is in
contrast with the typical nulling timescales of radio pulsars,
exceeding it by almost five orders of magnitude.

Cordes \& Shannon (2006) studied the possibility that
\psr\, might be in a binary system with a large asteroid, surrounded 
by a disk of small asteroids, moving in a $\sim 40$\,day eccentric
orbit around the pulsar. This scenario suggests that the interaction
between the pulsar magnetosphere and the asteroids can be responsible
both for the torque change and the intermittent radio activity of
\psr. Recently, Zhang, Gil \& Dyks (2007) have proposed that intermittent
pulsars are old isolated neutron stars which entered (or are about to
enter) the so called ``death valley'' (Chen \& Ruderman 1993), where
the polar cap voltage of the pulsar is not sufficient to power the
pair production process. However, this scenario still fails to account for the quasi-periodic pulsed emission of \psr.

Here we report on X--ray (\S\ref{chandra}) and optical
(\S\ref{optical}) observations of \psr\, taken with the {\it Chandra
X--ray Observatory} and the {\it Isaac Newton Telescope}. Furthermore,
in \S\ref{orbital} we derive constraints on possible orbital
parameters and companion mass by studying the radio timing
residuals. In \S\ref{binary} and \S\ref{discussion}, we discuss our
results and investigate in detail the possibility that  \psr\, resides in a binary system with a $\sim$35\,day orbital period and a low
mass companion star or a planet.

\section{X-ray observation}
\label{chandra}

The {\it Chandra} Advanced CCD Imaging Spectrometer (ACIS) observed
\psr on 2006 July 20th, for an on--source exposure time of 
$\sim$9.7\,ks. The target position was placed in the standard
back--illuminated ACIS S3 aimpoint, using the FAINT mode. We corrected
the astrometry for any processing offset and we cleaned the image for
hot pixels. Running CIAO {\tt celldetect} and {\tt wavedetect}
tools\footnote{for details refer to http://asc.harvard.edu/ciao/}, no
X-ray sources were detected in the whole ACIS--S3 CCD, while four
(unrelated) sources were detected in the other CCDs at $>5\sigma$
confidence level over the background. In particular, no photons were
detected in a $1^{\prime\prime}$ radius around the \psr\, radio
position (see Tab.\,1 for the RA and DEC). We then obtained a 99\%
upper limit on the source count rate of
$4.74\times10^{-4}$\,count\,s$^{-1}$ (Gehrels 1986).

Assuming a conservative absorption value of $N_{H} =
8.3\times10^{21}$\,cm$^{-2}$ (derived for the pulsar position from
Dickey \& Lockman~1990, hence considering the whole Galactic $N_{H}$
in that direction), and using the PIMMS calculator\footnote{for
details refer to http://heasarc.gsfc.nasa.gov/Tools/w3pimms.html}, we
derived a 99\% upper limit on the 0.3--10\,keV unabsorbed flux of
$7\times10^{-15}$\,erg\,s$^{-1}$\,cm$^{-2}$ or
$1.2\times10^{-14}$\,erg\,s$^{-1}$\,cm$^{-2}$, assuming a black body
(kT = 0.3\,keV) or a power law ($\Gamma = 2.5$) spectral
decomposition, respectively. At 4.6\,kpc these fluxes translate into a
99\% upper limit on the 0.3--10\,keV X-ray luminosity of
$1.7\times10^{31}$\,erg\,s$^{-1}$ and $3\times10^{31}$\,erg\,s$^{-1}$,
depending on the assumed X--ray spectrum. However, note that the
distance of 4.6\,kpc, as inferred from the pulsar DM (Kramer et
al.~2006) might well have an uncertainty of a factor of 2 due to a
$\sim$30\% uncertainty in the Galactic electron density model (Cordes
\& Lazio 2002). Hence, more conservative upper limits on the
luminosity would be 4 times larger than reported above.

\begin{center}
\begin{figure}
\centerline{
\psfig{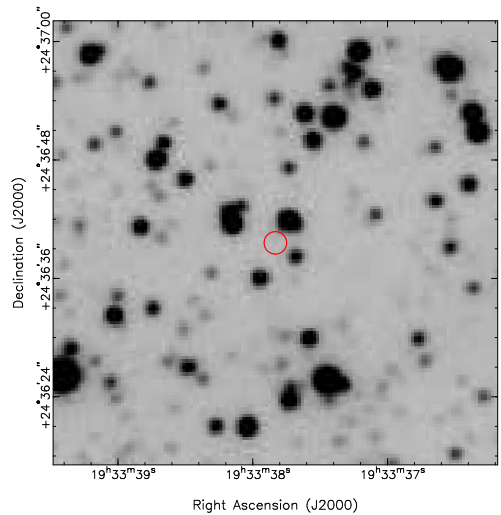}}
\caption{Finding-chart of the i$^{\prime}$ field of \psr. The circle is the uncertainty in the pulsar postion as derived from radio timing (see Tab\,1).}
\label{opt_obs}
\end{figure}
\end{center}

\section{Optical observations}
\label{optical}

We observed the field of \psr\, with the {\it Isaac Newton Telescope}
located at the Roque de Los Muchachos Observatory in La Palma, for an
exposure time of 10 minutes in three optical filters: g$^{\prime}$,
r$^{\prime}$ and i$^{\prime}$ (see Fig.~\ref{opt_obs} for the
finding-chart).  No optical counterpart was detected within the radio
pulsar position. We derived 5$\sigma$ upper limits on the optical
emission of \psr of g$^{\prime} > 22.6$, r$^{\prime} >22.4$ and
i$^{\prime} >22.2$ magnitudes. We inferred these optical upper limits
from the magnitudes of the faintest object detected at 5$\sigma$
confidence level in the same CCD as the pulsar position. For the astrometry we used an $11^{\prime} \times 11^{\prime}$
subsection of the g$^{\prime}$ image where we found 85 stars from the
USNO CCD Astrograph Calalogue (UCAC2; Zacharias et al.~2004). We
obtained an astrometric solution, fitting the zero--point position,
scale and position angle; the  final rms residuals were $0^{\prime\prime}.13$ in both in RA and DEC.

\begin{table}
\begin{center}
\begin{tabular}{lr}
\hline
Parameter & Value \\
\hline
RA (J2000) &  19:33:37.88(5) \\
DEC (J2000) &  24:36:40(1.5) \\
Epoch &   52281.296676 \\
$\nu$ (Hz) & 1.22896706877(3) \\
$\dot{\nu}$ ($10^{-15}$ s$^{-2}$) & $-12.1501(4)$ \\
$\ddot{\nu}$ ($10^{-25}$ s$^{-3}$) & $-2.0(2)$ \\
DM (cm$^{-3}$ pc)&   106.03(6) \\
\hline
\end{tabular}
\caption{Parameters derived from the radio timing analysis described in \S\ref{orbital}. Errors are at 1$\sigma$ confidence level.} 
\end{center}
\end{table}


We inferred the reddening in the direction of \psr\, from the $N_{H}$,
which gave an $A_{V}=4.64$ (Predehl \& Schmitt~1995). We then
converted this value into an estimate of the reddening in the filter
we actually used (Rieke \& Lebofsky~1985; Schlegel et al.~1998):
$A_{g^{\prime}} =4.94$, $A_{r^{\prime}}=3.68$ and
$A_{i^{\prime}}=2.83$. Considering a distance of 4.6\,kpc, we inferred
upper limits on the source absolute magnitude of: g$^{\prime} >4.35$,
r$^{\prime} >5.2$ and i$^{\prime} >6.1$. From our upper limits of the
absolute optical magnitudes, we could derive a rough range of stellar
types for a putative counterpart. From r$^{\prime} > 5.2$, which was
the most constraining limit, we could exclude all giant and
super-giant stars, and deriving a spectral type later than a G8
star. Note that having we assumed the absorption value of the whole Galactic
$N_{H}$: any (more realistic) lower $N_{H}$ values would result in stellar types much later than a G8.

\section{Limits on the orbital parameters from the radio timing}
\label{orbital}

If \psr\, happens to be in a binary system (e.g. as investigated in
detail in \S\,\ref{binary}), the motion of the pulsar
around the system's centre of mass should leave a periodic signature
in the remaining timing residuals. We searched for this signature in
radio timing data spanning almost 7 years (Kramer et al. 2006). We fit
the radio timing residuals with one spin frequency derivative and a
second derivative (values are reported in Tab.\,1). The latter was used
to remove a cubic term due to long term timing noise. In order to
improve the position information, we removed the remaining timing
noise by using the fit-wave method described by Hobbs et al. (2004),
with a minimum period of 2 years. The pre-fit-wave residuals were then
analysed for periodicities caused by low-mass companions (with masses
smaller then the pulsar mass, $M_c << M_p $), as described by Freire
et al. (2003). As Fig.\,2 shows, we detected, unsurprisingly, a faint
signal at a period of $\sim37$ days. However, we believe this signal
is consistent with a periodicity caused by the sampling of
time-of-arrivals forced upon us by the quasi-periodic on-phases of the
pulsar. We do not detect any other significant signal.

For a more general companion mass we also performed the following analysis 
of the timing residuals. The peak-to-peak amplitude of the remaining timing 
residuals was $\Delta t_{\rm res} \sim$2\,ms. From this we placed limits on the
corresponding orbital parameters by interpreting $\Delta t_{\rm res}$
as caused by a ``Roemer delay'', i.e.~the light-travel time across the
orbit (e.g.~Lorimer \& Kramer 2005). We then set:
\[
\Delta t_{\rm Roemer} = x \left[( \cos E - e)\sin\omega + 
\sin E \sqrt{1-e^2}\cos\omega\right] 
\]
\noindent
equal to $\Delta t_{\rm res}$. Here, $E$ is the eccentric anomaly, $e$
the eccentricity and $\omega$ the longitude of the
periastron. Furthermore, $x = a_{\rm p} \ \sin i /c$ is the projected
semi-major axis measured in light-seconds, which is a function of
$a_{\rm p} = a_{\rm R} \ M_{\rm c} / (M_{\rm p} + M_{\rm c})$, the
semi-major axis of the orbit\footnote{$i$ is the orbital
inclination angle, $c$ is the speed of light, and $a_{\rm R}$ is the size of
the relative orbit.}. 

On the other hand, the relative orbit $a_{\rm R}$, the orbital period
and masses are related according to Kepler's 3rd law:
\[
\frac{4\pi^2}{P_{\rm orb}^2} \left( \frac{a_{\rm R}}{c} \right)^3
= T_\odot (M_{\rm p} + M_{\rm c}),
\]
\noindent
with the masses measured in solar units and
$T_\odot=GM_\odot/c^3=4.925490947 \mu$s. Assuming $P_{\rm orb}
\sim 35$\,days and $M_{\rm p} \sim 1.4 M_\odot$ we could
derive an estimate for the orbital inclination angle as a function of
the companion mass in the following way: for a given companion mass
$M_{\rm c}$ we performed Monte-Carlo simulations drawing possible
values for $E$, $\omega$ and $e$ from uniform distributions over
$[0,2\pi]$ and $[0,1]$, respectively. Note that the distribution for
the eccentricity is not expected to be uniform (i.e.~it is likely to
be skewed to small eccentricities) but for our purposes this
assumption is sufficient as we were mostly interested in upper bounds
for the inclination angle.  For each companion mass, one million
Monte-Carlo runs were performed and the median and 95\% confidence
limits were computed (see Fig.~\ref{radio}: the grey shadowed region
is the allowed region at 95\% confidence level). Hence, for reasonable
companion masses the tight limit on a detectable periodicity in the
timing residuals implies rather small orbital inclination angles (see
also \S\ref{discussion}).

\section{Investigating the possible binary nature of \psr}
\label{binary}

In this section we investigate the possibility that the peculiar
characteristics of \psr\, can be explained by a low mass star or a
gaseous planet moving in an eccentric orbit around the pulsar (see
also \S\,\ref{orbital}). This causes a quasi-periodic
intermittent behavior, as well as the change in the spin-down
characteristics of the pulsar, in two different ways: 1) providing at
the periastron passage the additional plasma needed in the Kramer et
al. (2006) model, or 2) accreting material on the pulsar magnetosphere
and causing the transition between the radio pulsar and the propeller
regime (see for details Illarionov \& Sunyaev 1975; Stella et
al. 1986, 1994). In this section we discuss only this second
possibility. The physics involved in the first possibility has been
thoroughly investigated by Kramer et al.~(2006), although without
considering the wind of a companion as the source of the additional
magnetospheric plasma during the radio--on phase (see also
\S\,\ref{discussion}).

We define here three important \psr\, radii which will be used in the following.

\noindent
The magnetospheric radius:
\[
\, R_{{\rm m}} =  2\times10^{7} \dot{M}_{15}^{-2/7} B_9^{4/7} M_{1.4}^{-1/7} R_6^{12/7}
\simeq  \ 1.78\times10^{9} \dot{M}_{15}^{-2/7} \,{\rm cm} \ ,
\]
\noindent
the corotation radius:
\[
R_{{\rm cor}} = 0.12\times10^{8} M_{1.4}^{1/3} P_{10}^{2/3} \ 
\simeq \ 2.25\times10^{8} \, {\rm cm} \ ,
\]
\noindent
and the light cylinder radius:
\[ 
R_{{\rm lc}}= 0.46\times10^{8} P_{10}  \ 
\simeq  \ 3.74\times10^{9} \, {\rm cm} \ .
\]

\noindent
These three radii represent, respectively: i) the balance of the
gravitational force and the ram pressure of the infalling material,
ii) the centrifugal barrier for the infalling material due to the
pulsar rotation, and iii) the place where the pulsar magnetic field
lines open such that their tangential velocity do not exceed the speed
of light (Illarionov \& Sunyaev 1975; Ruderman \& Sutherland 1975). 

We define: $B_9 = B_{\rm ns}/10^{9}$\,G is the neutron star magnetic
field, $P_{10}$ is the spin period in units of 10\,ms, $\dot{M}_{15} =
\dot{M}/10^{15}$\,g~s$^{-1}$ is the mass inflow rate toward the neutron
star, and $M_{1.4} = M_{\rm ns}/1.4 M_{\odot}$ and $R_6 = R_{\rm ns} /
10^6$\,cm are the mass and the radius of the neutron star, hereafter
assumed $M_{1.4}= R_6 = 1$. Note that during the orbital motion, the
only variable radius is $R_{{\rm m}} \propto \dot{M}_{15}^{-2/7}$.

\begin{center}
\begin{figure}
\centerline{
\psfig{figure=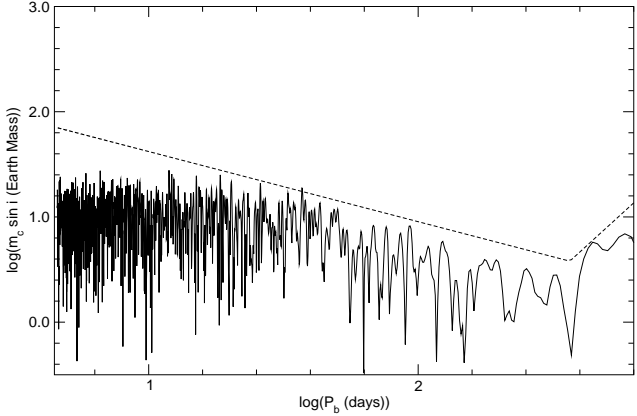,width=10cm,angle=-90}}
\caption{Lomb--Scale spectrum of \psr. Dashed curve indicates a 99.99\% 
confidence level. A periodicity is detected at $\sim37$\,days, but see
text for details.}
\label{radio}
\end{figure}
\end{center}


\subsection{Radio pulsar is on}
\label{on}

If the mass inflow rate from the putative companion star is very low,
as  might happen at apastron of an eccentric orbit, the
magnetosphere of the neutron star might be larger than the light
cylinder and corotation radii $(R_{{\rm m}} > R_{{\rm lc}} > R_{{\rm
cor}})$.

As a result, the centrifugal barrier is closed, which means that
incoming material from the companion star is prevented from falling
towards the neutron star magnetosphere or surface. In this case the
neutron star might behave as a radio pulsar. The pulsar radiation pressure dominates over the ram
pressure of the inflowing material, preventing the matter to penetrate
towards the neutron star (Illarionov \& Sunyaev 1975; Davis \& Pringle
1981; Stella et al.~1994; Campana et al.~1998). 

This occurs when the mass inflow rate ($\dot{M}$) is smaller than the
limiting value:

\[
\dot{M}_{15,{\rm radio-off}} = 5.4 \times 10^{-2} B_9^2 P_{10}^{-7/2} M_{1.4}^{-1/2} R_6^6 \ \simeq \ 0.075 \ , 
\]

\noindent
above which the radio pulsations would quench, even as rapidly as
$\sim$10\,s. 

 
It is worth noting that when $\dot{M} \sim \dot{M}_{15,{\rm radio-off}}$, the radio pulsar emission is not in equilibrium (Illarionov \& Sunyaev 1975; Shaham \& Tavani 1991), depending on the
stability of the mass inflow rate; sporadic variability of the radio--on duration of \psr\, might take place, as indeed observed (Kramer et al.~2006). Furthermore, during the radio--on phase, variations in the
pulsar DM as a function of the orbital phase are expected.  Unfortunately the observations were performed at one frequency, making impossible to put a meaningful limit on the DM variability (see also Kramer et al.~2006).

\subsection{Radio pulsar is off}
\label{off}

If the mass inflow rate starts to increase, e.g. approaching
periastron, we expect a correlated decrease of
the magnetospheric radius, which eventually becomes smaller than the light
cylinder radius $(R_{{\rm lc}} > R_{{\rm m}} > R_{{\rm cor}})$. This happens
when the mass inflow rate towards the neutron star becomes larger than
the limiting value $\dot{M}_{15,{\rm radio-off}}$; at this point the pulsar
radiation pressure is overcome by the ram pressure of the infalling
material quenching the radio pulsar mechanism. 
Different regimes are then allowed at this stage.  If $R_{{\rm
m}}$ remains larger than $R_{{\rm cor}}$, which means
$\dot{M}_{15,{\rm radio-off}} < \dot{M} < \dot{M}_{15,{\rm acc}}$, with

\[ \dot{M}_{15,{\rm acc}} = 5.97 \ B_9^2 P_{10}^{-7/3} M_{1.4}^{-5/3} R_6^6 \  
\simeq  \  1.4  \times10^{3},
\]
\noindent
then the magnetosphere of the neutron star still rotates in a
super-Keplerian motion, and the inflowing material might either
accumulate outside the magnetospheric boundary or be swept away by the
magnetospheric drag: this is called the ``propeller'' regime (Pringle \& Rees,
1972; Illaroniov \& Suniaev 1975; Davies \& Pringle 1981; Wang \&
Robertson 1983; Stella, White \& Rosner~1986).

Given the upper limits we derived for the possible companion star 
(see \S\,\ref{optical} and \S\,\ref{orbital}), there is only a very small
chance that the $\dot{M}$, due to the wind loss of the companion,
overcomes $\dot{M}_{15,{\rm acc}}$. We will then consider
hereafter only the possibility that the radio--off phase is driven by
the propeller regime, excluding the surface accretion scenario.

When the radio pulsations are quenched, the spin down behavior of the pulsar
is not driven anymore by the magnetic dipolar loss as it was
before.  What happens to the pulsar spin--down during the propeller regime is still
rather controversial, and requires detailed hydrodynamic
simulations (Romanova et al.~2003). Depending on which kind of
instability and shock takes place on the pulsar magnetosphere, the
spin--down rate might either increase further or be reduced by the
angular momentum transferred by the infalling material to the
magnetosphere.

In the \psr\, case, it is clear that the infalling material should provide
a certain rotational energy in order to make the pulsar slow down less
efficiently during the radio--off phase, changing the spin-down of the
neutron star by $\Delta \dot{\nu} = \dot{\nu}_{{\rm on}} - \dot{\nu}_{{\rm off}} = -
5.5\pm0.4\times10^{-15}$\, Hz\,s$^{-1}$, which converted in energy
corresponds to $\Delta \dot{E} \simeq 4 \pi^2 I \nu \Delta \dot{v} \ \simeq
2.6\times 10^{30} {\rm erg\,s}^{-1}$.

When the infalling matter reaches the neutron star magnetosphere, the source is expected to emit in the X--ray band with a minimum luminosity of:
\[
L_{{\rm radio-off}} = 7\times10^{31} B_9^2 P_{10}^{-9/2} M_{1.4}^{1/2} R_6^6 \ \simeq  \ 0.12\times10^{31} {\rm erg\,s}^{-1} ,
\]
\noindent
and a maximum luminosity related to the maximum wind
loss we expect from the inferred companion type. For a stellar type 
later than a G8, we expect the X--ray luminosity of the system in
the radio--off phase to be between $1.2\times10^{30} -
1.6\times10^{32}$\,erg\,s$^{-1}$.

After the periastron passage, the low-mass companion star moves towards
apastron, the magnetospheric boundary expands because of the
decreasing mass inflow rate, the centrifugal barrier closes and the
radio pulsar mechanism can resume again.

\begin{center}
\begin{figure}
\centerline{
\psfig{figure=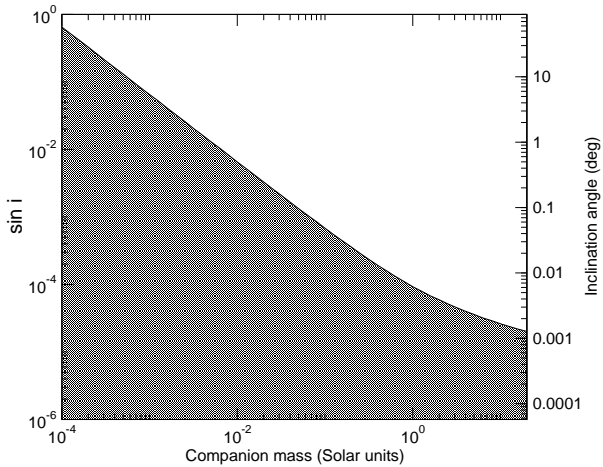,width=10cm,angle=-90}}
\caption{Limits on the orbital inclination angle as a function of 
the companion mass as derived from the pulsar radio timing. For each
companion mass, one million Monte-Carlo runs were performed, the
median and 95\% confidence limits were computed. The grey shadowed
region is the allowed region at 95\% confidence level.}
\label{radio}
\end{figure}
\end{center}


\section{Discussion}
\label{discussion}


The upper limits we derived from these X-ray and optical observations of
\psr\, during the radio--on phase, are not deep enough to detect the typical 
X--ray luminosity ($<10^{29}$\,erg~s$^{-1}$; Becker \& Tr\"umper 1997) 
and optical magnitude (V$>$28; Mignani et al.~2007) of an isolated pulsar 
of $\sim$1.6\,Myr, as \psr.  

On the other hand, if the pulsar is hosted in a binary system, we derived
tight  constraints on the putative companion and its orbital parameters. 
We constrained the companion to be a low mass star later than a G8 
type or a gaseous planet. 
For an orbital inclination of 0.1 degree the companion would
have a mass $M_{\rm c}<$0.01 M$_{\odot}$, while with a smaller
inclination of 0.01 degree, the companion mass is allowed to be
$M_{\rm c}<$0.1 M$_{\odot}$ (all at 95\% confidence level; see
Fig.\,3).

In the binary scenario (see \S\,\ref{binary}), the pulsar
intermittence can be explained by the transition between the radio
pulsar regime and the propeller regime during the orbital motion. The
X--ray upper limits we derived are unfortunately relative to the radio
pulsar dominated regime (note that the source was radio--on during the
observations), and not to the propeller regime when the source can be
bright in the X-ray due to accretion of matter onto the pulsar
magnetosphere. Our {\em Chandra} observation was in fact aimed at
detecting a possible X--ray emission from \psr\, during its radio--off
phase. Unfortunately, the relatively uncertain duration of both phases
(the radio--on phase rarely might happen to be longer than the average
duration), and the relatively large error ($\sim$ a few days) on the
putative periastron passage, made the observation be performed during
the radio--on phase of the pulsar. 

Note that the insufficient flux detection limit of current and past X--ray surveys and monitoring
programs (e.g. ROSAT and the RXTE/ASM), do not make the non
detection of an X--ray counterpart in previous X-ray surveys at all
constraining.

Although very successful in explaining the intermittence, the
quasi--periodicity and the torque variations of \psr, this binary
scenario still presents several unsolved issues. One issue of this scenario, and to some extent also to the ``asteroid'' model (Cordes \& Shannon 2006), is the very small orbital
inclinations required by the radio timing studies (\S\,\ref{orbital}
and Fig.\,3). For a random distribution of inclination angles, the
probability of observing a system at an angle less than a value $i_0$,
is p$(i<i_0) = 1 - \cos{i_0}$. Considering the favorable case of an angle of 15$^{\circ}$ (a gaseous planet; see Fig. 3), the probability of observing such a system is p$(i<15^{\circ})= 3.4$\%.  A chance of 3.4\% is not very high, although still worth to be considered.

Another issue concerns the accretion rate from the companion. Can the
accretion from the low wind loss of such a small mass star or gaseous
planet be high enough to switch the pulsar from the radio emitting
regime to the propeller regime?  Typical wind rates for
e.g. a K0 star are insufficient to produce the limiting value of
$\dot{M}_{15,{\rm radio-off}}$ by $\sim$2 orders of magnitude. This
problem might be partially (but not totally) alleviated by taking into
account the irradiation process (Podsiadlowski 1991; D'Antona \& Ergma
1993). In particular, a low mass star irradiated by the radio pulsar
is expected to expand even if the radiation bath is not
particularly extreme, and the wind of the star to increase substantially. 
However, the irradiation process cannot be dominant on the stellar wind from a G8 to K0 star, mainly because of the relatively low rotational energy of \psr\, with respect to
the surface temperature of the companion star ($L_{irr} = f
\dot{E}_{\rm rot} (R_{s}/2a_{\rm p})^2$, where $R_{s}$ is the
companion star radius). For lower mass stars or considering a gaseous planet 
orbiting around the pulsar, this process might have instead a substantial role, and might produce
strong winds towards the pulsar while at periastron.

If the weak stellar wind from the low-mass companion star is insufficient
to drive the radio pulsar in the propeller regime, it is instead strong 
enough to provide the amount of plasma needed to produce the
spin--down change in the scenario proposed by Kramer et
al. (2006). Note that in this picture, at the periastron passage we
expect the radio--on phase, while in the propeller scenario it
is expected at the apastron of the orbit.  However, while
the propeller scenario naturally explains the quenching of the radio
emission, in this picture the radio quenching still remains puzzling: 
we would in fact expect to see
the pulsar at all the time, although with two different spin--down
behavior.

\section*{Acknowledgments}

We acknowledge the director of the Chandra X-ray Observatory, Harvey
Tananbaum, for promptly according us an observation of \psr\, through
his Director Discretionary Time, and for useful comments on this
draft. We also thank the Chandra team for the efficiency during the
scheduling process. We are grateful to the referee, S.~Ransom, for his
valuable suggestions which largely improved our work.  NR
acknowledges support from an NWO Veni Fellowship and thanks
F. Verbunt, M. Burgay, D. Russell, and the ATNF--Epping pulsar group,
for useful comments and discussions.

\label{lastpage}

\end{document}